# A Review on Orbital Angular Momentum With the Approach of Using in Fifth-Generation Mobile Communications

Seyed Ali Khorasani
Department of Electrical Engineering
Sharif University of Technology
Tehran, Iran
ali.khorasani@ee.sharif.edu

*Abstract*— In this paper, different generations of mobile communication have been concisely mentioned. The need for advanced antenna systems capable of sending and receiving massive data is felt in the fifth generation of mobile communication. The beamforming method and multi-input multi-output systems (MIMO) are the proposed solutions to increase the channel capacity of the communication network. Orbital angular momentum (OAM), an inherent feature of electromagnetic waves, is a suitable solution to increase channel capacity. This feature will increase the channel capacity by producing orthogonal modes. Using antenna arrays is an effective way to produce these modes. The results of FEKO simulations show the capability of this method.

## I. INTRODUCTION

The first generation of mobile communication networks was introduced in 1979. The defects of this technology did not cause it to be abandoned so that by 1990, more than 20 million users were using it. Due to challenges such as poor coverage, low voice quality, lack of inter-operator roaming services, and most importantly lack of encryption of conversations, the research into the second generation (2G) systems began. The second generation of communication was first accomplished in 1991 under the GSM standard. For the first time, conversations were encrypted and digital voice calls were much clear than before. That was the first time people sent short messages, video messages and multimedia messages to each other. Despite the slow speed of communication, this generation can be considered a revolution in the communication systems. The third generation of communication was used in 2001. The speed increase to more than 4 times of the second generation, and the possibility of making video calls and international roaming were among the features of this generation. The fourth generation had the capability of fast access to web pages, high-quality video calls and a speed of over 1 gigabit per second (1 Gbps) [1, 2]. The main point was the equipment required for this technology. This opportunity led to significant progress, especially in the design of mobile phones.

## II. FIFTH GENERATION MOBILE COMMUNICATIONS

By comparing the fourth and fifth generations of mobile communication systems, it can be seen that the data transfer speed has improved 10 to 100 times, which means the data transfer speed will reach almost 10 gigabits per second (10 Gbps). 5G technology achieves a latency of less than 1 millisecond. To figure out this time, it is enough to know that it takes about 250 milliseconds for a human to react. It is possible to perform systems such as intelligent vehicles, remote surgeries, and smart power networks that require a platform with almost zero latency by this feature. Increasing the network capacity is one of the advances of the fifth generation compared to others, which enables another concept called the Internet of Things. Advanced antenna systems are suitable to use in current and future telecommunication networks [3]. Today is the best time to replace old systems with advanced antennas. The advantages of these antennas can be listed as follows: Proper performance in connection with satellites and ground users, economic justification, cost reduction of digital processing in beamforming technology, and in MIMO [4, 5, 6]. This is suitable for operators who want to improve their coverage and increase their capacity. The fifth-generation network standards should meet the requirements such as increasing capacity, being suitable for short-range communication with the least dissipations, controlling path dissipations limits, and simultaneous multi-user connection with the least latency. Millimeter waves are strongly influenced by atmospheric conditions and raindrops and may even be absorbed by oxygen molecules and water vapor. The solution to this problem is to reduce the range covered by each antenna; since the distance increases, the dissipations increase exponentially. Delay and throughput are two main factors in advanced communication systems. Antennas play an essential role in the fifth generation of mobile communication, so adaptable and dynamic antennas in different conditions are required. By designing a new configuration, it is possible to provide a mechanism for the antenna to select its frequency band. One way to increase the channel capacity is to use MIMO array antennas [7]. The above system can simultaneously support several communication channels, each of which works separately. The dimensions of the antennas are enlarged at frequencies less than 3 GHz and cause limitations for integrating a large number of antennas. The antenna's dimensions become smaller in the spectrum of millimeter waves, which helps multi-channel transmission. Integrating broadband antennas increases the total output power of the system. It is also necessary to use high-gain antennas to deal with dissipation caused by working at high frequencies. According to a general rule, increasing the dimensions of the antenna leads to an increase in the antenna's directivity. In fact, by arraying an antenna, the dimensions of the antenna increase. One of problems of array antennas is the presence of unwanted extra grating lobes in the horizontal and vertical planes of the radiation pattern, which correct placement will be a suitable solution to solve this.

## III. Orbital Angular Momentum of Electromagnetic Waves

In electrodynamics, fields can also carry momentum, unlike gravity which only bodies can. Electromagnetic fields can store energy which causes the conservation of momentum in electrodynamics. The electromagnetic force on a density of charge is:

$$\vec{F} = \int_V (\vec{E} + \vec{v} \times \vec{B})\rho d\tau = \int_V (\rho\vec{E} + \vec{J} \times \vec{B})d\tau \quad (1)$$

We introduce Maxwell's stress tensor to describe the force per unit volume easier.

$$T_{ij} = \epsilon_0 \left(E_i E_j - \frac{1}{2}\delta_{ij}E^2\right) + \frac{1}{\mu_0}(B_i B_j - \frac{1}{2}\delta_{ij}B^2) \quad (2)$$

The above subscripts are related to the components of the Cartesian coordinates, so the tensor has 9 components. Using the above tensor, the force per unit volume is:

$$\vec{f} = \nabla \cdot \overleftrightarrow{T} - \epsilon_0 \mu_0 \frac{\partial \vec{S}}{\partial t} \quad (3)$$

By integrating on both sides and using the divergence theorem, the total force on the charge density is:

$$\vec{F} = \oint_S \overleftrightarrow{T} \cdot \vec{da} - \epsilon_0 \mu_0 \frac{d}{dt} \int_V \vec{S} d\tau \quad (4)$$

By comparing the above expression with Newton's second law, an expression similar to Poynting's theorem is obtained in which the first integral represents the momentum stored in electromagnetic fields and the second integral represents the momentum passing through the surface per unit of time. In general, the linear and angular momentum of electromagnetic fields are as follows:

$$\vec{P} = \epsilon_0 (\vec{E} \times \vec{B}) \quad (5)$$
$$\vec{L} = \vec{r} \times \vec{P} = \epsilon_0 [\vec{r} \times (\vec{E} \times \vec{B})] \quad (6)$$

The above expressions show that even in static fields if the product of $\vec{E} \times \vec{B}$ is non-zero, they can still have linear and angular momentum [8]. As has been said before, electromagnetic waves carry energy and momentum. Momentum itself divides into linear momentum and angular momentum. Angular momentum, which consists of spin angular momentum (SAM) and orbital angular momentum (OAM), is related to wave polarization using these two. The connection between them can be seen in how electrons move around the nucleus of an atom. The momentum caused by the moving of the electrons around the nucleus is equal to the orbital angular momentum, and the one caused by the rotation of the electrons around themselves is equal to the spin angular momentum. For the first time, the idea of using orbital angular momentum was proposed along with optical vortices. In an optical vortex, constant phase planes of the electric and magnetic fields are moving spirally in the direction of propagation. One of the characteristics of the vortex is the topological charge number, which indicates the twists of light around the axis in a wavelength. The larger this number is, the more twists there are. Theoretically, the angular momentum carried by the optical vortex has an infinite number of eigenstates defined in the infinite-dimensional Hilbert space [9]. That is why orbital angular momentum has numerous applications in telecommunications [10, 11, 12, 13]. If the orbital angular momentum of photons is fully used to carry and share information, the capacity of a photon increases dramatically. As a result, the transmission capacity of single-mode and single-wavelength fibers also increases. The capabilities of using orbital angular momentum are not limited to the visible light spectrum. For example, we can point out to perform this concept in wireless communication and radio frequencies lower than visible light. Applications of orbital angular momentum in wireless communications and underwater acoustic communications (which are very difficult due to high dissipations, multiple propagation paths, small bandwidths, etc.) have created new research fields. The magnitude of the vortex wave field in the center of the propagation axis is zero; therefore, it is displayed in black. The spiral structure of these waves is defined by exponential functions $exp(i\theta l)$, where θ is the transverse angle and l is the topological charge (orbital angular momentum mode). One of the most significant features of orbital angular momentum beams is their orthogonality. As a result, the inner product of two beams with different topological charge numbers will be:

$$\int_0^{2\pi} e^{i\ell_1 \theta} (e^{i\ell_2 \theta})^* d\theta = \begin{cases} 0 & \ell_1 \neq \ell_2 \\ 2\pi & \ell_1 = \ell_2 \end{cases} \quad (7)$$

Due to this, waves with different topological charge numbers can be considered separate information sets in data transmission paths. These new extra dimensions lead to an increase in system capacity. One of the most fundamental ways to produce these waves is using the Spiral Phase Plate (Fig. 1) [14, 15]. In summary, it should be said that these plates create a spiral shape by making various phase differences on different parts of the wavefront. Using metamaterial reflectors is another way to produce these waves. We can consider antenna arrays as another way of generating waves based on orbital angular momentum. The uniform circular array antennas are the most common for producing electromagnetic vortex waves [16, 17].

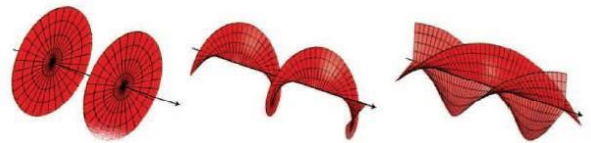

*Fig. 1. Topological Charge Number (Increasing from Left to Right) [15]*

This array consists of *N* components that are uniformly placed on the circumference of a circle. Each component is fed by its corresponding signal, which is similar in magnitude to others but has a phase difference of $\Delta\varphi = \frac{2\pi l}{N}$ with its adjacent component. *l* is the orbital angular momentum mode (topological charge) which means that the phase difference of the whole array is $2\pi l$. These waves can be separated by using the orthogonality property explained earlier [18]. In the receiver, phase recovery is the main key to determining the different modes. One of the ways to recover waves is to use reverse spiral plates.

IV. SIMULATION

In this section, the phase pattern of the electric field of the antenna arrays is extracted using FEKO software. At first, current point sources at the frequency of 10 GHz have been used as excitation sources. It can be seen that the phase pattern of the electric field caused by a single-point current source has a zero mode. The phase pattern of higher modes is obtained by increasing the number of sources. Fig. 2 shows the electric field phase pattern in 0, 1, 2, 3, 4, 5, and 6 modes. By considering how the number of sources changes with generated modes, a relationship can be reached to calculate the minimum number of sources to produce waves based on the orbital angular momentum.

$$N \geq 2|l| + 1 \quad (8)$$

*l* and *N* are the topological charge number and the minimum number of array elements, respectively. Table 1. shows the minimum number of sources required to achieve the desired mode. Now, current point sources replace by dipole antennas. The length of this antenna is chosen in a way so that its electric field, based on the mean square error, is at most 10% different from the field caused by current point sources.

The simulation has been done at two frequencies of 3 GHz and 86 GHz using this antenna array. Fig. 3 shows the phase pattern of this antenna array at the frequency of 3 GHz, with the length of each antenna equal to $10\lambda$. As can be seen, the phase pattern of the electric field in desired modes was obtained by using this antenna array. By studying the pattern of the electric field magnitude in the electric and magnetic field planes, the big difference between them is determined. The reason is that the point sources emit all their power, while dipole antennas reflect most of it. The S-parameter diagram of the array confirms this.

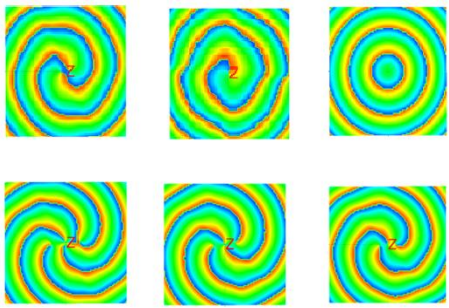

Fig. 2. Phase Pattern of Electric Field due to Point Source Array

Table 1. Minimum Number of Point Sources to Produce Desired Mode

| Desired Mode | Min No. of Point Source |
|---|---|
| 1 | 3 |
| 2 | 5 |
| 3 | 7 |
| 4 | 9 |
| 5 | 11 |

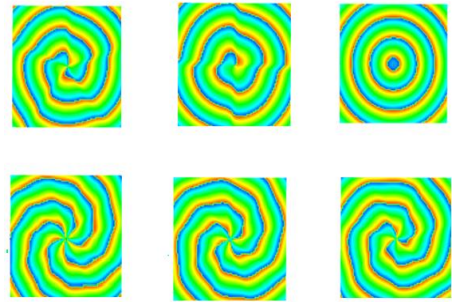

Fig. 3. Phase Pattern of Electric Field due to Array of Antenna

The electric field of the antenna arrays, based on the magnitude of the angular momentum, is proportional to a sum in the space in which various factors are effective. Angular momentum mode, the distance of each element from the computing point, and the spatial phase difference between the array elements are some of these factors [19].

$$E(r,\varphi,z) \propto \sum_{n=1}^{N} a_n \cdot \frac{e^{-ikR_n}}{R_n} exp(-il\phi_n) \quad (9)$$

$$R_n = \sqrt{a^2 + z^2 + r^2 - 2ar\cos(\varphi - \varphi_n)} \quad (10)$$

From the theory of antennas, it is determined that the electric field caused by an array is always proportional to an expression called the array coefficient, which has its shape and depends on what type of array is used, whether linear, planar, or circular. In the following, the coefficient of a circular array will be compared with the above expression.

$$E(r,\vartheta,\varphi) = \sum_{n=1}^{N} a_n \cdot \frac{e^{-ikR_n}}{R_n} \quad (11)$$

R is the distance of the nth element from the calculation point, and $a_n$ is the excitation coefficient of the nth element. By looking at the two expressions of the electric field, it can be seen that the electric field of an ordinary circular array; is a specific array for producing an electric field with different modes $l = 0$. In fact, by adding the phase caused by producing different modes to an ordinary array, we can obtain one with the ability to generate different modes. According to the above expression, by increasing the number of array elements, the magnitude of the field also increases. Generally, increasing the dimensions of the antenna leads to an increase in the magnitude and directivity of the field radiation pattern. Actually, by increasing the number of elements, the dimensions of the antenna increase. Until now, array elements were regularly arranged on the circumference of a circle with a radius equal to the wavelength. Now, these elements are placed in the environment of a smart mobile phone, and their results are checked. The elements of the array are placed once on the center of the sides of the largest surface of the phone (irregular) and once again on the vertices of a square to the width of the phone (regular) (fig. 4). These simulations are performed at 3 GHz and 86 GHz frequencies.

The phase distribution pattern of the electric field of the mobile phone antenna array in regular and irregular modes is shown in Fig. 5 and Fig. 6. By studying these two figures, it is clear that the phase pattern in regular mode has a better condition. In the following, the S-parameter diagram is drawn in regular array mode at 3 GHz frequency (fig. 7). Based on a comparison of this diagram, it can be concluded that the level of reflection of waves between each input is extremely low (non-diagonal regions of the matrix of dispersion characteristics), while the level of reflection between each input and itself (diagonal regions of the matrix of dispersion characteristics) is also almost zero decibels. As a result, almost all of the energy reaching the input is reflected, and very little of it is radiated. Due to this reason, dipole antenna arrays have lower electric field amplitude levels than arrays with (hypothetical) point sources.

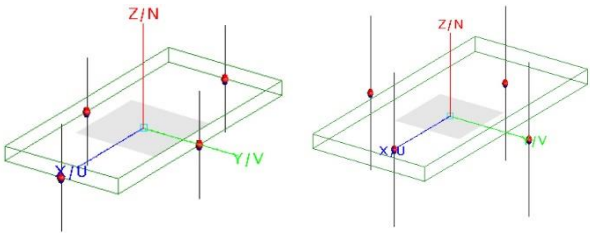

*Fig. 4. Placement of Antennas, Regular Array and Irregular Array*

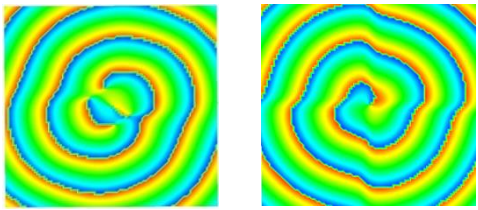

*Fig. 5. Phase Pattern of Electrical Field of Smart Phone Array, Regular and Irregular, 3GHz*

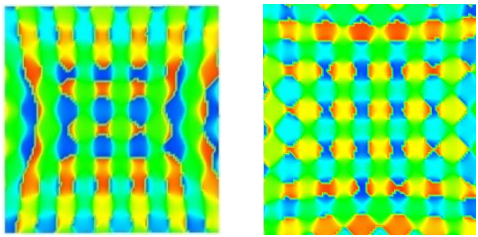

*Fig. 6. Phase Pattern of Electrical Field of Smart Phone Array, Regular and Irregular, 86GHz*

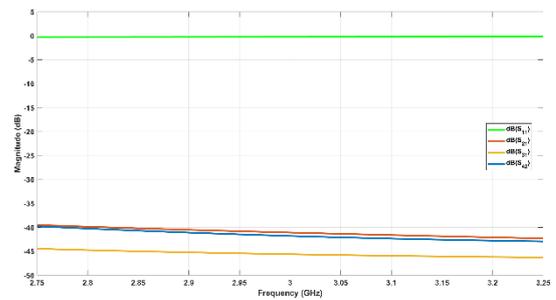

*Fig. 7 S-parameter Diagram of Smart Phone Array, Regular, 3GHz*

It is also observed that the regular cell array generally produces a better electric field phase pattern than the irregular cell array. All of the arrays above are not ideal radiation arrays since most of the power reaching the antennas is not radiated. It is not possible to achieve the desired electric field phase pattern at 86 GHz because of the very small antenna length compared to the array elements' distance. The simulation of circles of point sources results in a favorable pattern even at high frequencies. This is because with a frequency change, the distance between elements also changes. In the matrix of the S-parameter of arrays, the magnitude of the non-diagonal elements decreases dramatically with increasing frequency. In addition to increasing electric field amplitude level, increasing the number of array elements improves the phase distribution pattern.

CONCLUSION

Fifth generation mobile communication requires advanced antenna systems. Increasing the capacity of the communication channel is one of the main concerns in this generation. Using the characteristic of orbital angular momentum as an inherent characteristic of electromagnetic waves can solve this issue. Antenna arrays based on Orbital Angular Momentum (OAM) can be used to generate waves of desirable modes, which are orthogonal to one another. As a result of this feature, the fifth generation of mobile communication channels have a greater capacity.

ACKNOWLEDGMENT

The Author would like to express his very great appreciation to Mr. Seyed Amir Hossein Khorasani for his valuable and constructive suggestions during the planning and development of this research work. His willingness to give his time so generously has been greatly appreciated.